\documentclass[showpacs,twocolumn,superscriptaddress,nofootinbib]{revtex4-1}

\usepackage{hyperref}
\hypersetup{
  colorlinks=true,         
  linkcolor=blue,         
  citecolor=cyan,         
}

\usepackage{graphicx}
\usepackage{dcolumn}
\usepackage{bm}
\usepackage{color}
\usepackage{footmisc}

\usepackage{amsmath}
\usepackage{amssymb}


\begin{document}

\title{Efficiency of magnetic 
Penrose process\\ in higher dimensional Myers-Perry black hole spacetimes}

\author{Sanjar Shaymatov}
\email{sanjar@astrin.uz}
\affiliation{Institute for Theoretical Physics and Cosmology, Zhejiang University of Technology, Hangzhou 310023, China}
\affiliation{Institute of Fundamental and Applied Research, National Research University TIIAME, Kori Niyoziy 39, Tashkent 100000, Uzbekistan}

\date{\today}
\begin{abstract}

In this paper, we consider a well-established magnetic Penrose process (MPP) and bring out its impact on the efficiency of energy extraction from higher dimensional (i.e., $D>4$) black holes. We derive the field equations of motion and determine the expressions for the energy efficiency of energy extraction for the case of higher dimensional black holes. We also examine the efficiency of energy extraction from black holes with $(n-1)$ and $n$ rotations. We demonstrate that black holes with $(n-1)$ rotations has only one horizon, resulting in infinitely large energy efficiency even without MPP. On the other hand, for black holes with $n$ rotations in $D>4$, the energy efficiency is not infinitely large, but the efficiency can be significantly enhanced by MPP. This enhancement allows for arbitrarily large energy efficiency. We find that the efficiency of energy extraction can exceed over $>100\%$ for $D=5,6$ and $D=7,8$ dimensions. Interestingly, for rotation parameters near the extremal value, the energy efficiency remains above $100 \%$ in $D=7,8$ compared to $D=5,6$. MPP can eventually make higher dimensional black holes more efficient even with $n$ rotations.

\end{abstract}
\pacs{%
} \maketitle

\section{Introduction \label{Sec:introduction}}

Black holes are believed to form as a consequence of the gravitational collapse of massive stars and their existence is regarded as a generic result of Einstein’s general theory of relativity. Thus, they have so far been regarded as the most fascinating compact objects due to their remarkable nature and aspects. However, black holes had been considered candidates due to the absence of direct detection. After the detection of gravitational waves as a consequence of two black hole mergers \cite{Abbott16a,Abbott16b} and the first image of the supermassive black hole located in the center of the galaxy M87 by BlackHoleCam and the Event Horizon Telescope (EHT) collaborations \cite{Akiyama19L1,Akiyama19L6}, these modern observations opened up new avenues to verify the existence of black holes in nature and to provide potential explanations for some high energy X-ray sources \cite{King01ApJ}, active galactic nuclei (AGNs) \cite{Peterson:97book}, and gamma-ray bursts \cite{Meszaros06} that give rise to an enormous amount of released energy in general relativity (GR). Also, these observations can be considered as direct tests to probe the remarkable nature of their geometry in the close vicinity of the black hole horizon as well as highly energetic astrophysical events associated with energies of the order of $10^{42}-10^{47}\:\rm{erg/s}$ ~\cite{Fender04mnrs,Auchettl17ApJ,IceCube17b}. Therefore, astrophysical black holes have been considered as sources of highly energetic astrophysical phenomena and as key points in explaining these extremely powerful events \cite{McHardy06,Woosley93ApJ,Preparata98,Popham98,Rappaport05}. 
Energy extraction mechanisms come into play and become increasingly important in explaining these highly energetic powerful astrophysical phenomena by extracting the rotational energy from rapidly rotating black holes. 

The Penrose process (PP) \cite{Penrose:1969pc} for energy extraction mechanism was proposed as a potential explanation for highly energetic astrophysical events around black holes. Penrose formulated this process theoretically and showed that it is possible to extract the rotational energy of a rapidly rotating black hole. This was formulated by taking advantage of the ergosphere, which exists in the region between the black hole's horizon and the static limit that is bounded by the surface from the outside. To extract the energy from the black hole, there must exist an ergosphere where an infalling particle splits into two pieces so that one falls into the black hole, while the other escapes to infinity with larger energy than the original one. With this, the energy of the escaping particle can be extracted from black hole, thus resulting in the black hole slowing down. However, there are limitations to the Penrose process, such as the low cross section, which makes it difficult to extract more energy from the black hole. Bardeen et al. \cite{Bardeen72} and later Wald \cite{Wald74ApJ} independently showed that for PP to be efficient, the original incident particle or relative velocity of the two splitting pieces has to be relativistic. Otherwise, the energy extraction can never be larger. Despite these limitations, the Penrose process has been considered and extended to a large variety of contexts, including Kerr-Newman Taub-NUT spacetime \cite{Abdujabbarov11}, rotating regular black holes~\cite{Toshmatov:2014qja}, collisional Penrose process with spinning particles~\cite{Okabayashi20}, and axially symmetric magnetized Reissner-Nordström black holes \cite{Shaymatov22b}.

It is worth noting, however, that PP has since been considered highly theoretical and not directly observed. There can exist extreme conditions in the surrounding environment of rotating black holes that make the process complicated to analyze in detail. Therefore, this mechanism may not provide valuable insights into the behavior of black holes associated with the source of highly energetic astrophysical phenomena in the universe. In this context, the original PP may not be efficient enough to address the most powerful energy sources, such as the outflows coming out from active galactic nuclei (AGNs) and highly energetic quasars in the universe. Later, in 1985 Wagh, Dhurandhar and Dadhich \cite{Bhat85} reformulated the original PP and proposed a new mechanism as the magnetic Penrose process (MPP). In this process, the weak magnetic field existing in the close environment surrounding a black hole plays a crucial role in providing the energy required for an escaping particle to ride on a negative energy orbit, allowing it to overcome the constraint velocity of the original Penrose process. The magnetc Penrose process is analogous to the Blandford-Znajeck mechanism \cite{Blandford1977}, which utilizes the electromagnetic processes to extract the rotational energy of a rotating Kerr black hole. This addresses the impact of a purely magnetic field on the released energy from AGNs (see, for example \cite{McKinney07}). The magnetic Penrose process ~\cite{Bhat85,Parthasarathy86} has been considered an efficient mechanism for demonstrating the effect of a purely magnetic field on the energy extraction process over the years (see, for example \cite{Wagh89,Alic12ApJ,Moesta12ApJ,Nozawa05,Dadhich18mnras,Tursunov:2019oiq,Tursunov20ApJ,Shaymatov22b}). It should also be noted that other theoretical explanations for highly energetic objects have been proposed, such as magnetic reconnection \cite{Comisso21MR} and superradiance processes \cite{Brito20Book}, leading to energy extraction from black holes. This has also sparked increased research activity, with various authors investigating these processes in different scenarios \cite{Kumar_Jha23,Rahmani20,Khodadi21PLB,Khodadi22,Liu22ApJ,Khodadi22MR,Wang22,Khodadi23MR,Shaymatov23MRP}. 

As mentioned, black holes are very fascinating gravitational objects in the universe. There is also an interesting solution that describes axially symmetric five dimensional black hole spacetime in the supergravity of the Einstein-Maxwell equation \cite{Chong05a}. An interesting point to note is that a rotating black hole in higher dimensions ($D>4$) would have more than one rotation axis \cite{Myers-Perry86}. Therefore, five or six dimensional black holes can have two rotation parameters with the two axes. Following \cite{Myers-Perry86,Chong05a}, there are investigations in which the energetic properties of a higher dimensional rotating/charged black hole were demonstrated in Refs.~\cite{Prabhu10}, addressing the energy extraction process. This was all done in the absence of a magnetic field in the environment surrounding the black hole. There is also an extensive analysis that has been done regarding the linear and non-linear accretion process for higher dimensional Myers–Perry (MP) rotating black holes \cite{An18,Shaymatov19a,Shaymatov20a,Shaymatov20b}.  

In this paper, we investigate higher dimensional MP rotating black holes in the presence of an external magnetic field. We explore MPP for black holes with $(n-1)$ and $n$ rotations in odd $D=2n+1$ and even $D=2n+2$ dimensions, based on the spacetime geometry introduced by Myers and Perry \cite{Myers-Perry86}. The point to be noted is that rotation in higher dimensions ($D>4$) has two aspects: more than one rotation and the other through $(n-1)$ and $n$ rotations with various horizon structures. This is a remarkable and distinguishing feature of these higher dimensional black holes, which does not exist in the $D=4$ dimensional analogue.  The main objective of this study is to analyze the efficiency of energy extraction in these scenarios and generalize all higher dimensional black holes with $(n-1)$ and $n$ rotation cases, highlighting the impact of the MPP and providing valuable insights into their energetic properties. We find that the energy extraction efficiency is significantly different for black holes with $(n-1)$ rotations compared to those with $n$ rotations. Surprisingly, black holes with even PP exhibit a higher efficiency in extracting rotational energy compared to those with $n$ rotations. However, we demonstrate that it is still possible for black holes with $n$ rotations to efficiently extract their rotational energy using MPP.  


The paper is organized as follows: In Sec.~\ref{Sec:metric}, we discuss higher dimensional ($D>4$) MP rotating black hole spacetime, which can be further used to formulate MPP in both odd and even dimensions. In Sec.~\ref{Sec:MF_PD}, we proceed to review the magnetic field and charged particle dynamics in the close surrounding environment of MP black hole. In Sec.~\ref{Sec:MPP1}, we consider MPP and derive its energy efficiency expressions for higher dimensional MP black holes together with the magnetic field part in odd $D=2n+1$ and even $D=2n+2$ dimensions. In Sec.~\ref{Sec:MPP2}, we discuss the efficiency of energy extraction from black holes with $(n-1)$ and $n$ rotations and analyze its results together with further comparisons in the given $D>4$ dimension. Finally, we end up with a conclusion in Sec.~\ref{Sec:Conclusion}. Throughout we use the following system of units in which $G=c=1$.

\section{Higher dimensional Myers–Perry black hole spacetime } \label{Sec:metric}

The metric describing the well-known higher dimensional Myers-Perry (MP) rotating black holes in Boyer-Lindquist coordinates is given by \cite{Myers-Perry86}
\begin{eqnarray}\label{Eq:D}
ds^2&=&-dt^2+r^2d\beta^2 + \sum_{i=1}^{n}(r^2+a^2_{i})\left(d\mu_{i}^2+\mu_{i}^2d\phi^2_{i}\right)\nonumber\\&+&\frac{\mu r}{\Pi F}\left(dt +\sum_{i=1}^{n}a_{i}\mu_{i}^2d\phi_{i}\right)^2 +\frac{\Pi F}{\Delta}dr^2\, ,
\end{eqnarray}
with
\begin{eqnarray}\label{Eq:D1}\label{eq:F}
F &=& 1-\sum_{i=1}^{n}\frac{a_{i}^2\mu_{i}^2} {r^2+a_{i}^2}\, , \\
\label{eq:Pi}
\Pi &=&\prod_{i=1}^{n}(r^2+a_i^2)
\, , \\
\label{eq:delta}
\Delta &=& \Pi -\mu r^{2n-D+3}\, ,
\end{eqnarray}
where $\mu$ and $a_{i}$ refer to black hole mass and rotation parameters, respectively, while $n=[(D-1)/2],\, [(D-2)/2]$ to the maximum number of rotations black hole can have in given $D=2n+1,\,2n+2$ dimensions. Here, $\mu_i$ and $\beta$ for $D=2n+2, 2n+1$ are given by the following relations~\cite{Myers2011}, 
\begin{eqnarray}\label{Eq:2n+2}
\sum_{i=1}^{n} \mu_i^2 + \beta^2 &=& 1\, ,\\
\label{Eq:2n+1}
\sum_{i=1}^{n} \mu_i^2 &=& 1\, .
\end{eqnarray}
It should be noted here that Eq.~(\ref{Eq:2n+1}) and the metric for $D=2n+1$ are always satisfied in the limit of $\beta=0$ . To be more informative, we define $\mu_i$ and $\beta$ by the following expressions with the direction cosines; i.e. for $D=2n+2$  
\begin{eqnarray}\label{Eq:con_laws}
\begin{cases}
\mu_1=\sin\theta_1\, \\
\mu_2=\cos\theta_1\sin\theta_2\, \\
\makebox[4em]{\dotfill}\\
\mu_n=\cos\theta_1\cos\theta_2\cdots\sin\theta_n\, \\
\beta=\cos\theta_1\cos\theta_2\cdots\cos\theta_n\, , 
\end{cases}
\end{eqnarray}
and $D=2n+1$
\begin{eqnarray}
\begin{cases}
\mu_1=\sin\theta_1\, \\
\mu_2=\cos\theta_1\, \\
\makebox[4em]{\dotfill}\\
\mu_{n-1}=\cos\theta_1\cos\theta_2\cdots\sin\theta_{n-1}\, \\
\mu_{n}=\cos\theta_1\cos\theta_2\cdots\cos\theta_{n-1}\, .  
\end{cases}
\end{eqnarray}
For example, $\mu_i$ and $\beta$ for $D=5,6$ dimensions take the following forms as  
\begin{eqnarray}
\mu_1=\sin\theta_1\, \mbox{~~and~~}\mu_2=\cos\theta_1\, ,
\end{eqnarray}
and 
\begin{eqnarray}
\mu_1=\sin\theta_1\, ,\,\,  
\mu_2=\cos\theta_1\sin\theta_2\, \,\,\mbox{and} \,\,\,
\beta=\cos\theta_1\cos\theta_2\, .\nonumber\\
\end{eqnarray}

The horizon equation is given by 
\begin{eqnarray}
\Delta=\Pi -\mu r^{2n-D+3}=0\, ,
\end{eqnarray}
which solves to give black hole horizon in odd $D=2n+1$ and even $D=2n+2$ dimensions, depending on the number $n$ of rotation parameters, which will be further discussed (see for example \cite{Aliev04MP,Shaymatov19a,Shaymatov21a}). It is to be emphasized that black hole can be endowed with more rotations $n=[(D-1)/2],\,[(D-2)/2]$ in higher $D=2n+1,\,2n+2$ dimensions; e.g. $n=2,\,3$ for $D=5,6$ and $D=7,8$ dimensions, respectively. 

Another interesting point to be noted is that, due to stationary and axial symmetry the higher dimensional MP rotating black hole can permit having more than two Killing vectors \cite{Myers-Perry86}, i.e. 
\begin{align}
    \xi_{(t)} = \frac{\partial}{\partial t} \ , \quad \xi_{(\phi_1)}= \frac{\partial}{\partial \phi_1}\ , \quad\cdots\, , \quad \xi_{(\phi_n)}=\frac{\partial}{\partial \phi_n} \, .
\end{align}
 This is an interesting aspect of all higher dimensional MP black holes. It is to be emphasized that, for example, in $D=5,\,6$ dimensions there exist three Killing vectors, i.e., $\xi_{(t)}$ represents a stationary, while the rest two $\xi_{(\phi_1)}$ and $\xi_{(\phi_2)}$ exhibit an axisymmetry of $D=5,\,6$ dimensional black hole spacetime. There does therefore exist three the conserved quantities correspondingly, such as the energy and two angular momenta for a test particle with mass $m$ due to aforementioned Killing vectors that are also sufficient to construct electromagnetic $n$-vector potential for the Maxwell test field in the higher dimensional MP rotating black hole spacetime. It should also be noted that $n$-velocity of the zero angular momentum observer (ZAMO) in higher dimensional MP rotting black hole spacetime can be defined by \cite{Aliev04MP} 
 \begin{align} 
 u^\mu = \alpha\left(\xi_{(t)}^\mu +\Omega_{\phi_1} \xi_{(\phi_1)}^\mu + \cdots+\Omega_{\phi_n} \xi_{(\phi_n)}^\mu\right)\, , 
 \end{align}
which is perpendicular to the surface with $t = const$ that implies $r=const$ and $\theta_i=const$, i.e., $u^r=0$ and $u^{\theta_{i}}=0$. Here, $\alpha$ are referred to as a normalization constant which can be determined by imposing the normalization condition $g_{\mu\nu}u^\mu u^\nu =-1$. We now turn to describe the static limit surface where the time-like Killing vector $\xi_{(t)}$ of the metric turns out to be null; i.e. $g_{tt}=0$  that solves to give $r_{st}$ implicitly. For example, $r_{st}$ for $D=5,\,6$ dimensions will respectively read as follows \cite{Prabhu10}: 
\begin{eqnarray}
    r_{st} = \sqrt{\mu - \kappa}\, ,
\end{eqnarray}
and 
\begin{eqnarray}
    r_{st} = \frac{2^{1/3}\left(9 \mu+\sqrt{81 \mu^2+12 \kappa^3}\right)^{2/3} - 2\,3^{1/3}\kappa}{6^{2/3}\left(9 \mu+\sqrt{81 \mu^2+12 \kappa^3}\right)^{1/3}}\, ,
\end{eqnarray}
where $\kappa=a_1^2\cos^2\theta +a_2^2\sin^2\theta$.
It is worth noting that for the energy extraction through PP there must exist the ergosphere which occurs in the region between the black hole horizon $r_{+}$ and the static limit $r_{st}$ bounded by the surface from outside, i.e., $r_{+} < r < r_{st}$. Note that the black hole's ergosphere takes a different $r_{st}$ in the given $D$ dimension accordingly. This is one advantage of having ergosphere for the energy extraction via either PP \cite{Penrose:1969pc} or MPP \cite{Bhat85}.

\section{The magnetic field and the charged particle dynamics}\label{Sec:MF_PD}

The aim of this section is to consider the magnetic field and the charged particle motion around the higher dimensional MP rotating black hole spacetimes described by the line element of Eq.~(\ref{Eq:D}). To this end, black hole is supposed to be placed in a uniformly distributed magnetic field. With this regard, black hole can experience the magnetic field being uniform at large distances and being weak enough for any change in the background geometry as a test field (i.e., which is of order $B_1\sim 10^{8}~\rm{G}$ and $B_2\sim 10^{4}~\rm{G}$ for stellar mass and supermassive black holes, respectively ~\cite{Piotrovich10,Baczko16,Daly:APJ:2019:}). It is to be emphasized that the magnetic field was estimated to be of order $B\sim 33.1 \pm 0.9~\rm{G}$ in the corona as suggested by the observational analysis for binary black hole system $V404$ Cygni~\cite{Dallilar2018}. Very recently, EHT collaborations reported that observations at 230 GHz were able to image the polarized synchrotron radiation in the vicinity of the supermassive black hole sitting at the center of the M87 galaxy, probing the magnetic field structure in the environment surrounding the black hole and suggesting that the average magnetic field strength is of order $B\sim 1 - 30~\rm{G}$ in the emission region~\cite{MF:2021ApJ,Narayan2021ApJ}. It is worth noting that, although the magnetic field is small it can strongly influence the geodesics of charged particles.
Modeling the charged particle motion around a black hole endowed with an external uniform magnetic field has been developed by Wald and since been widely used as the well-known formalism \cite{Wald1974ApJ}. The point to be noted here is that Killing vectors in a vacuum spacetime come into play to construct electromagnetic potential for the Maxwell test field~\cite{Papapetrou:1966zz}. Hence, the magnetic field solution can be obtained as a test field in the surrounding environment of background spacetime. It is to be emphasized that the magnetic field, although small, influences the motion of charged particles drastically than the gravity because of the Lorentz force. It does therefore contribute to changing the motion of charged particles in the curved background spacetime.  There has been a large amount of work devoted to the study of
of the magnetic field impact on charged particle motion and and on different astrophysical processes in the close surrounding environment of black holes~\cite[see, e.g.][]{Wald:1974np,Aliev02,Frolov10,Shaymatov20egb,Tursunov16,Shaymatov22a,Shaymatov21pdu,Shaymatov21c,Hussain17,Shaymatov22c,Shaymatov23GRG}. 

Unlike Kerr black hole in four dimensions, the higher $D>4$ dimensional MP metric, as mentioned, admits more than two Killing vectors; i.e. the stationary $\xi_{(t)}$ and axial ones $\xi_{({\phi_{1}\,\cdots\,\phi_{n}})}$ of the spacetime Eq.~(\ref{Eq:D}). It was also discussed by \cite{Aliev:2005npa} addressing the electromagnetic field existing in the surrounding environment of five dimensional black hole spacetime.  Afterwards, Wald's method \cite{Wald1974ApJ,Aliev:2005npa} permits one to write the Maxwell field equation for the vector potential $A^{\mu}$ in the MP spacetime,  and it is given by \cite{Aliev2002MNRAS}
\begin{align} \label{Eq:maxwell_field}
    A_{;\nu}^{\mu;\nu}-R_\nu^\mu A^\nu = 0 \, ,
\end{align}
which can have the same form for Killing vector $\xi$ as
\begin{align}\label{Eq:killing_field}
    \xi_{;\nu}^{\mu;\nu}-R_\nu^\mu \xi^\nu = 0 \, ,
\end{align}
where $R_{\mu\nu}$ refers to the Ricci tensor. It is clearly seen from Eq.~(\ref{Eq:killing_field}) the second term associated with Ricci tensor gets vanished in the vacuum case, thereby taking the simpler form $\Box \xi^\mu=0$. This implies the same form as $\Box A^\mu=0$ of the Maxwell equations for the vector potential. Taking all together, $n$-vector potential of the electromagnetic field in the $n$ dimensional black hole spacetime can be written as \cite{Aliev:2005npa,Kunz:2005nm}  
\begin{eqnarray}\label{Eq.EM_pot}
A^{\mu}=C_0\, \xi^{\mu}_{(t)}+ \sum_{i=1}^{n}C_i\, \xi^{\mu}_{(\phi_{i})}\, ,
\end{eqnarray}
where we refer to the uniform external magnetic field strength as $ B_{i}$ in relation to $\phi_{i}$ rotation planes. Here, $C_0$ and $C_i$ are usually referred to as arbitrary parameters as integration constants that stem from the field properties (see for example \cite{Aliev04MP}). We note that the $n$-vector potential involves a uniformly magnetic field and a Coulomb-type component which can be induced by the rotation of black hole. It happens because the magnetic field is more likely to be distorted because of black hole rotation, thereby giving rise to the induced charge which can remain present regardless of absence of black hole electric charge (i.e. $Q = 0$). However, for our purpose and further analysis we will omit this component so that the time component of $A^{\mu}$ reads as 
\begin{eqnarray}\label{Eq.EM_pot}
    A^t=\sum_{i=1}^{n}\frac{B_i}{n}\left(\xi_{(\phi_i)}^t+a_{i}\xi_{(t)}^t\right)\, .
\end{eqnarray}

We now examine the charged particle motion around the higher dimensional MP rotating black hole placed in an external uniform magnetic field. As mentioned, we assume that the magnetic field considered here is uniform at large distances and weak enough for any change in the background geometry as a test field. With this in view, we consider the Hamiltonian for a charged particle as 
\cite{Misner73}
\begin{align}\label{HJ}
H=\frac{1}{2}g^{\mu\nu}\left(\pi_{\mu}-qA_\mu\right)\left(\pi_{\nu} -q A_\nu\right) \, ,
\end{align}
where $\pi_{\mu}$ and $A_{\mu}$ are referred as the canonical momentum and the vector potential of electromagnetic field, which are written by the following relation 
\begin{eqnarray}
p^{\mu}=g^{\mu\nu}\left(\pi_{\nu}-qA_{\nu}\right)\, . 
\end{eqnarray}
It is to be emphasized that, according to the spacetime metric (\ref{Eq:D}) the canonical momentum facilitates, for example, three constants of motion which denote the particle's energy and angular momenta relative to $\phi_{1}$ and $\phi_{2}$ planes in $D=5,\,6$ dimensions. Further, we also discuss $n+3$ conserved quantities. In the following, we write the Hamilton's equations of particle motion as 
\begin{eqnarray} 
\label{Eq:eqh1}
  \frac{dx^\alpha}{d\lambda} = \frac{\partial H}{\partial \pi_\alpha}\,   \mbox{~~and~~}
  \frac{d\pi_\alpha}{d\lambda} = - \frac{\partial H}{\partial x^\alpha}\, ,
\end{eqnarray}
where we have denoted $\lambda=\tau/m$ as affine parameter with the proper time $\tau$ for time-like geodesics. Eq.~(\ref{Eq:eqh1}) then allows one to define the abovementioned constants of motion for timelike geodesics, and they are given by 
\begin{eqnarray}
\begin{cases}
\label{Eq:en} \pi_t-qA_{t}=
g_{tt}p^{t} + \sum_{i=1}^{n}g_{t\phi_{i}}p^{\phi_{i}}
\, ,\\
\pi_{\phi_{1}}-qA_{\phi_{1}}= g_{\phi_{1} t}p^{t} + \sum_{i=1}^{n}
g_{\phi_{1}\phi_{i}}p^{\phi_{i}}
\, , \\
\makebox[4em]{\dotfill}\\
\pi_{\phi_{n}}-qA_{\phi_{n}}= g_{\phi_{n} t}p^{t} +\sum_{i=1}^{n}
g_{\phi_{n}\phi_{i}}p^{\phi_{i}}
\, . 
\end{cases}
\end{eqnarray}
where the metric functions stem from Eq.~(\ref{Eq:D}). Taking altogether, the action for $D=\begin{cases}2n+1\\2n+2\end{cases}$ can be written accordingly to the Hamilton-Jacobi equation as
\begin{eqnarray}
    {\cal S}=\frac{1}{2}m^2\tau-Et+\sum_{i=1}^{n}L_{\phi_{i}}\phi_{i}+{\cal S}_r+\begin{cases}\sum_{i=1}^{n-1}{\cal S}_{\theta_{i}}\\\sum_{i=1}^{n}{\cal S}_{\theta_{i}} \end{cases}\, , 
\end{eqnarray}
 where the quantities $E \equiv -\pi_t$ 
  and $L_{\phi_n} \equiv \pi_{\phi_{n}}$ refer to the constants of motion, i.e., test particle's energy and $n$-angular momenta relative to $\phi_{n}$ axes, whereas ${\cal S}_{\theta_{_i}}$ and ${\cal S}_r$ represent the radial and angular functions of $r$ and $\theta$, respectively. It should be noted here that the system has $n+3$ independent constants of motion, namely, we have described $n+2$, such as $E$, $L_{\phi_{1}\,\cdots\,\phi_{n}}$ and $m^2$ (where $m$ is the mass of particle). $(n+3)$th one is described by the latitudinal motion of particles. We will not focus on it when considering motion to either the equatorial plane or polar plane in the higher $D>4$ dimensional MP spactime (i.e., $\theta_1=\theta_2\cdots\theta_n=\pi/2\, , 0$).  

In particular for $D=5,\,6$ dimensions (i.e., with two rotation axes $\phi_1$ and $\phi_2$) Eqs.~(\ref{Eq:en}) permit to derive the $n$-velocity components required to define the effective potential for radial motion of test particles as
\begin{eqnarray}
\Gamma &p^{t}&=
-\bigg[\left(E+qA_{t}\right) (g_{\phi_1\phi_2}^2 -
g_{\phi_1\phi_1} g_{\phi_2\phi_2}) \nonumber\\&-& \left(L_{\phi_2}-qA_{\phi_2}\right)g_{\phi_1\phi_1} g_{
t\phi_2} \nonumber\\&+&
\big(\left(L_{\phi_2}-qA_{\phi_2}\right) g_{t\phi_1} + \left(L_{\phi_1}-qA_{\phi_1}\right)
g_{t\phi_2}\big)g_{\phi_1\phi_2}\nonumber\\&-& \left(L_{\phi_1}-qA_{\phi_1}\right) g_{\phi_2\phi_2}g_{ t\phi_1}
\bigg],\,\\
\Gamma &p^{\phi_1}&=
-\bigg[\left(E+qA_{t}\right) \left(g_{\phi_2\phi_2}
g_{t\phi_1} -  g_{\phi_2\phi_1} g_{t\phi_2}\right) \nonumber\\&+& \big(\left(L_{\phi_2}-qA_{\phi_2}\right)
g_{t\phi_1}  - \left(L_{\phi_1}-qA_{\phi_1}\right) g_{t\phi_2}\big) g_{t\psi}\nonumber\\&-& 
\big(\left(L_{\phi_2}-qA_{\phi_2}\right)
g_{\phi_2\phi_1} -\left(L_{\phi_1}-qA_{\phi_1}\right) g_{\phi_2\phi_2}\big) g_{tt}\bigg],\,\nonumber\\\\
\Gamma &p^{\phi_2}&=
-\bigg[ \left(E+qA_{t}\right)( g_{\phi_1\phi_1}
g_{t\phi_2}- g_{\phi_2\phi_1} g_{t\phi_1}) \nonumber\\&-& \big(\left(L_{\phi_2}-qA_{\phi_2}\right) g_{t\phi_1}
- \left(L_{\phi_1}-qA_{\phi_1}\right) g_{t\phi_2}\big) g_{t\phi_1}\nonumber\\&+&  
\big(\left(L_{\phi_2}-qA_{\phi_2}\right)
g_{\phi_1\phi_1}  - \left(L_{\phi_1}-qA_{\phi_1}\right) g_{\phi_2\phi_1}\big) g_{tt}\bigg]\, ,\nonumber\\
\end{eqnarray}
where we have defined $\Gamma$ and electromagnetic potentials as  
\begin{eqnarray}
\Gamma &=&g_{\phi_2\phi_2} g_{t\phi_1}^2 - 2 g_{\phi_2\phi_1}
g_{t\phi_1} g_{t\phi_2} + g_{\phi_1\phi_1} g_{t\phi_2}^2 +
g_{\phi_2\phi_1}^2 g_{tt} \nonumber\\&-& g_{\phi_1\phi_1} g_{\phi_2\phi_2}
g_{tt}\, .
\end{eqnarray}
By plugging the above velocity components into the normalization condition, $g_{\mu\nu}p^{\mu}p^{\nu}=-m^2$, the effective potential for timelike radial motion of the charged particle in the equatorial plane (i.e. $\theta_1=\theta_2=\pi/2$) can be generally obtained as follows:  \cite{Dadhich22b,Dadhich22a}
\begin{eqnarray}\label{Eq:eff_pot}
V_{eff}(r)&=&-\frac{q}{m}A_{t}+\omega \left(
\mathcal{L}_{\phi}-\frac{q}{m}A_{\phi}\right)\nonumber\\&+&\sqrt{\frac{\Delta}{g_{\phi\phi}\,r^{2(n-1)}}\left(\frac{
\left(\mathcal{L}_{\phi}-\frac{q}{m}A_{\phi}\right)^2}{g_{\phi\phi}}+1\right)}\, ,
\end{eqnarray}
where $\mathcal{L}_{\phi}$ (for further analysis we shall for simplicity use $\phi_1\to \phi$) represents angular momentum of the charged particle, while $\omega = -g_{t\phi}/g_{\phi\phi}$ refers to the frame dragging angular velocity. Here, we note that $n$ is referred to as maximum number of rotation parameters in higher dimensions. It is to be emphasized that the effective potential is a valuable tool to gain a deeper understanding in relation to how test particles move in the close environment of black holes. Determining the minimum value of the effective potential allows to define the location of the innermost stable circular orbits (ISCOs) referring to the closest possible path for the particle to be on circular orbit without being fallen into the black hole or kept on coming out to space. With this in view, occurrence of no bound orbits around black holes in higher dimensions was discussed in Refs.~\cite{Dadhich22b,Dadhich22a}.  

We further need to determine the bounds of the charged particle's angular velocity. To that the circular motion of the charged particle is required together with $r=const$ and $\theta_{i}=const$. Afterwards, we can examine with the case ${\bf u}\sim {\bf \xi}_{(t)}+\sum_{i=1}^{n}\Omega_{\phi_{i}} {\bf \xi}_{(\phi_{i})}$, where $\Omega_{\phi_{i}}=d\phi_{i}/dt=u^{\phi_{i}}/u^{t}$ 
are referred to as the angular velocities measured by far away observer. As mentioned, we note that we restrict the motion of charged particle to the equatorial plane; i.e. $\theta_1=\theta_2\cdots\theta_n=\pi/2$. It implies the condition which satisfies the bound of angular velocity as $\Omega_{-}<\Omega<\Omega_{+}$ for the timelike vector ${\bf u}$, and it is given by 
\begin{eqnarray}
  \Omega_{\phi}^{\pm} =\frac{-g_{t\phi}\pm \sqrt{(g_{t\phi})^{2}-g_{tt}g_{\phi\phi}}}{g_{\phi\phi}}\, .
\end{eqnarray}
The above equation reduces to $\Omega_{\phi}^{+}=0$ and $\Omega_{\phi}^{-}=-2\omega$ at the static surface; i.e. $g_{tt}=0$. It is however valuable to note that the limiting values of $\Omega=\Omega_{\phi}^{\pm}$ refers to the photon motion. We then write the $n$-momentum for the circular motion of the charged particle with $r=const$ as 
\begin{eqnarray}\label{Eq:4-mom}
\pi_{\pm}=p^{t}(1,0,0,\Omega_{\phi_{1}}^{\pm},\cdots,\Omega_{\phi_{n}}^{\pm})\, ,
\end{eqnarray}
which gives the following equation for the time like circular motion ($\upsilon_{(r,\theta_{i})}=0$) at the equatorial plane (i.e. $\theta_1=\theta_2\cdots\theta_n=\pi/2$)
\begin{eqnarray}\label{Eq:W0}
\left(g_{\phi\phi}\pi_t^2+g_{t\phi}^2\right)\Omega^2&+&2g_{t\phi}\left(\pi_t^2+g_{tt}\right)\Omega\nonumber\\&+& g_{tt}\left(\pi_t^2+g_{tt}\right)=0\, ,
\end{eqnarray}
with $\pi_t=-\left(\mathcal{E}+qA_{t}/m\right)$. 
It is then obvious from the above equation that the angular velocity for the circular orbit of charged particle can be obtained by~\cite{Parthasarathy86,Nozawa05,Shaymatov22b}
\begin{eqnarray}\label{Eq:31}
\Omega_{\phi}=\frac{-g_{t\phi}\left(\pi_t^2+g_{tt}\right)+\sqrt{\left(\pi_t^2+g_{tt}\right)\left(g_{t\phi}^2-g_{tt}g_{\phi\phi}\right)\pi_t^2}}{g_{\phi\phi}\pi_t^2+g_{t\phi}^2}\, .\nonumber\\
\end{eqnarray}

Following expressions for angular velocity and charged particle motion we intend to investigate the energy extraction from black hole through MPP in the next section.

\section{Magnetic Penrose process in higher dimensions} \label{Sec:MPP1}

We now focus on MPP, according to which a falling neutral particle within the ergosphere of black hole is divided into two parts, that is one continues to fall into the black hole, while the other keeps on coming outward to space \cite{Penrose:1969pc}. The point to be noted here is that the escaping part can take out some energy from a rapidly rotating black hole, thereby resulting in making black hole slow down. Here, the energy the escaping particle carried away depends on the magnetic field strength, and thus the MPP becomes increasingly important in driving out the black hole rotational energy. We now turn to examine a scenario, in which a massive particle that has energy with $E_1\geq 1$ and keeps on falling inwards splits into two charged parts together with energies $E_{2}$ and $E_{3}$ in the ergosphere. As stated by MPP, the one keeps on falling inwards black hole with energy $E_{2}<0$, whereas the other flings away into space with positive energy $E_{3}>0$. Following to conservation principles for energy and angular momentum, this phenomenon at the splitting point can be written as 
\begin{align}\label{Con}
& E_1=E_2+E_3\, ,\quad L_{\phi\,1}=L_{\phi\,2}+L_{\phi\,3}\, ,\quad  \\  
& m_1 = m_2 + m_3\, , \quad  q_1=q_2+q_3\, .
\end{align}
It should be noted that the condition given in the above satisfies $E_{2}<0$ and $E_{3}\gg E_1$ as stated by the MPP. Interestingly, the particle that escapes from black hole acquires $E_{2}$ exceeding over the energy of incident particle, thereby extracting the rotational energy of black hole. Also, one can depict the momentum as \cite{Bhat85,Nozawa05}
\begin{eqnarray}\label{Eq:con_law}
m_1u_1^{\mu}&=& m_2u_2^{\mu}+m_3u_3^{\mu}\, ,
\end{eqnarray}
where we keep only $\phi$ component of $n$-velocity which is defined by $u^{\phi}=\Omega_{\phi} u^{t}=-\Omega_{\phi} F/B$. Consequently, one can rewrite Eq.~(\ref{Eq:con_law}) as follows:  
\begin{eqnarray}
\Omega_{\phi\,1}m_1F_{1}B_2B_3=\Omega_{\phi\,2}m_2F_{2}B_3B_1+\Omega_{\phi\,3}m_3F_{3}B_2B_1\, ,\nonumber\\
\end{eqnarray}
with the following notations $F_{i}=\mathcal{E}_{i} +q_{i}A_{t}/m_{i}$ and $B_{i}=g_{tt}+\Omega_{\phi\,i} g_{t\phi}$. In doing so, the above equation can be simplified as follows:  
\begin{eqnarray}
\frac{E_3+q_3A_{t}}{E_1+q_1A_{t}}=\left(\frac{\Omega_{\phi\,1}B_2-\Omega_{\phi\,2}B_1}{\Omega_{\phi\,3}B_2-\Omega_{\phi\,2}B_3}\right)\frac{B_3}{B_1}\, .
\end{eqnarray}
From the above equation the energy the escaping particle carried away can be obtained as 
\begin{eqnarray}\label{Eq:En3}
E_3=\chi\left(E_1+q_1A_{t}\right)-q_3A_{t}\, ,
\end{eqnarray}
where $\chi$ is given by  
\begin{eqnarray}\label{Eq:chi}
\chi=\left(\frac{\Omega_{\phi\,1}-\Omega_{\phi\,2}}{\Omega_{\phi\,3}-\Omega_{\phi\,2}}\right)\frac{B_3}{B_1}\, . 
\end{eqnarray}
To be more informative we have defined $\Omega_{\phi\,i}$ as follows:  
\begin{eqnarray}
\Omega_{\phi\,1}= \Omega_{\phi}\, , \mbox{~~} \Omega_{\phi\,2}=\Omega_{\phi}^{-}\, \mbox{~~and~~} \Omega_{\phi\,3}=\Omega_{\phi}^{+}\, .
\end{eqnarray}
To gain a deeper understanding we now examine the energy efficiency in relation to the maximum amount of energy extracted through the MPP. Let us then write the simpler form of the energy efficiency as 
\begin{eqnarray}
\eta= \frac{\vert E_2\vert}{E_1}=\frac{E_3-E_1}{E_1}\, .
\end{eqnarray}
Eq.~(\ref{Eq:En3}) permits the energy efficiency of the MPP at the horizon to be depicted as
\begin{eqnarray}
\eta= \chi-1+\frac{q_3A_t}{m_1\pi_{t1}+q_1A_t}-\frac{ q_1A_t}{m_1\pi_{t1}+q_1A_t}\,\chi\, ,
\end{eqnarray}
where $\chi$ is given by
\begin{eqnarray}
\chi&=&\left(\frac{\Omega_{\phi}-\Omega_{\phi}^{-}}{\Omega_{\phi}^{+}-\Omega_{\phi}^{-}}\right)\left(\frac{g_{tt}+\Omega_{\phi}^{+}g_{t\phi}}{g_{tt}+\Omega_{\phi}\,g_{t\phi}}\right)\nonumber\\
&=&\frac{g_{\phi\phi}\left(\sqrt{g_{tt}+\pi_t^2}+1\right)+g^2_{t\phi}}{2g_{\phi\phi}\sqrt{g_{tt}+\pi_t^2}}\, .
\end{eqnarray}

We now turn to analyse energy efficiency pertaining to the maximum energy extracted from black hole due to the radiation of incident particle split into two parts in the ergoregion. We shall then consider the possible case in which $q_1=0$ can be taken, i.e, $q_2+q_3=0$ as per the conservation principle for two split particles which can have either positive and negative. For further analysis we for simplicity take $q_3=q=-q_2$. With this regard, the energy efficiency takes the following form at the splitting point as 
\begin{eqnarray}
\eta= \left(\frac{\Omega_{\phi}-\Omega_{\phi}^{-}}{\Omega_{\phi}^{+}-\Omega_{\phi}^{-}}\right)\left(\frac{g_{tt}+\Omega_{\phi}^{+}g_{t\phi}}{g_{tt}+\Omega_{\phi}\,g_{t\phi}}\right)-1-\frac{q\,A_t}{E_1}\, .
\end{eqnarray}
\begin{figure*}	
\includegraphics[width=\columnwidth]{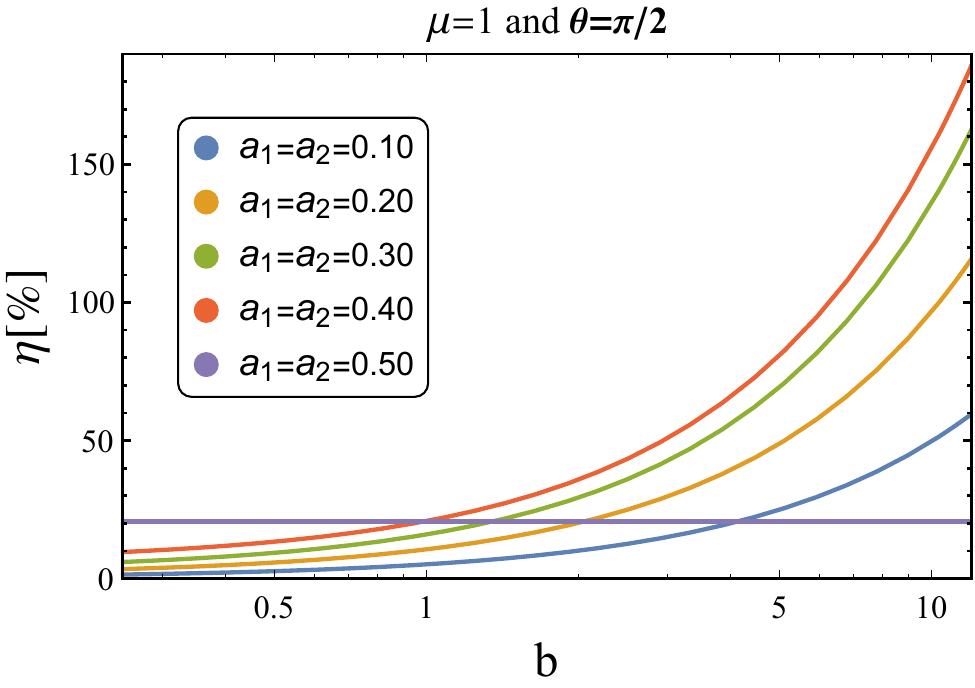}
\includegraphics[width=\columnwidth]{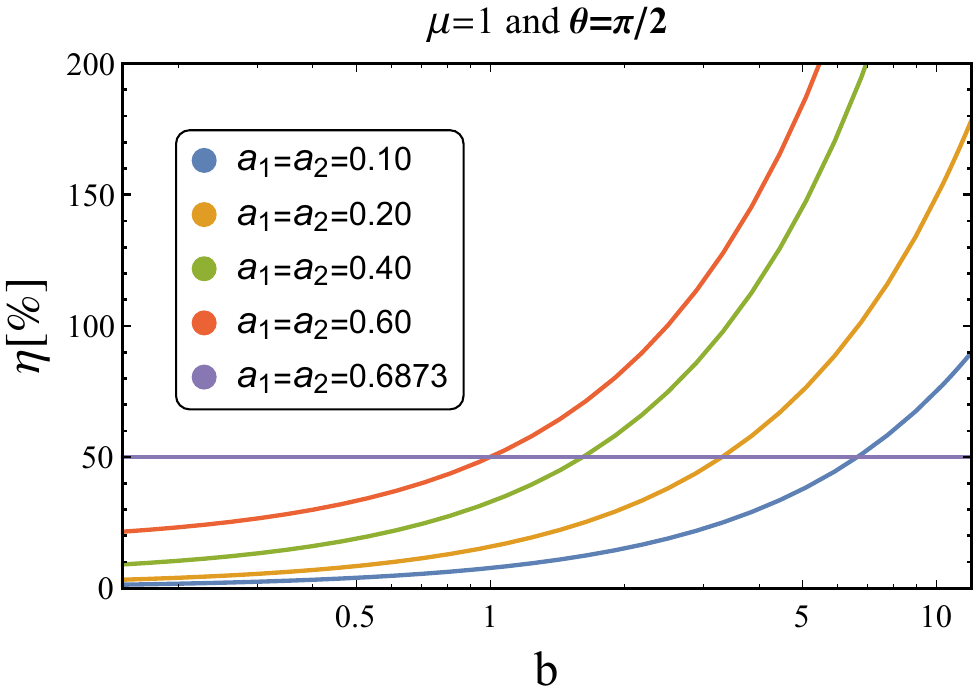}

\includegraphics[width=\columnwidth]{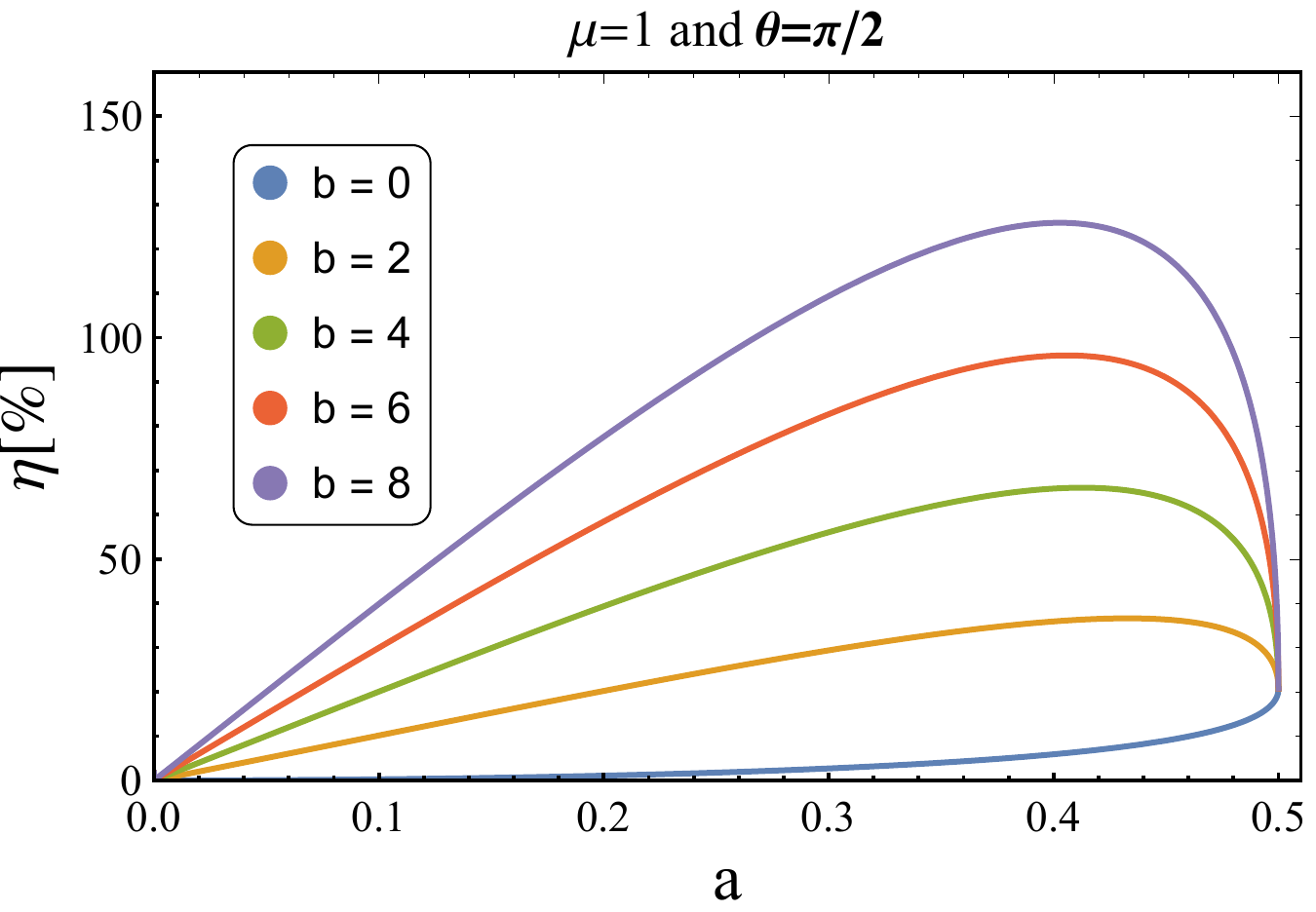}
\includegraphics[width=\columnwidth]{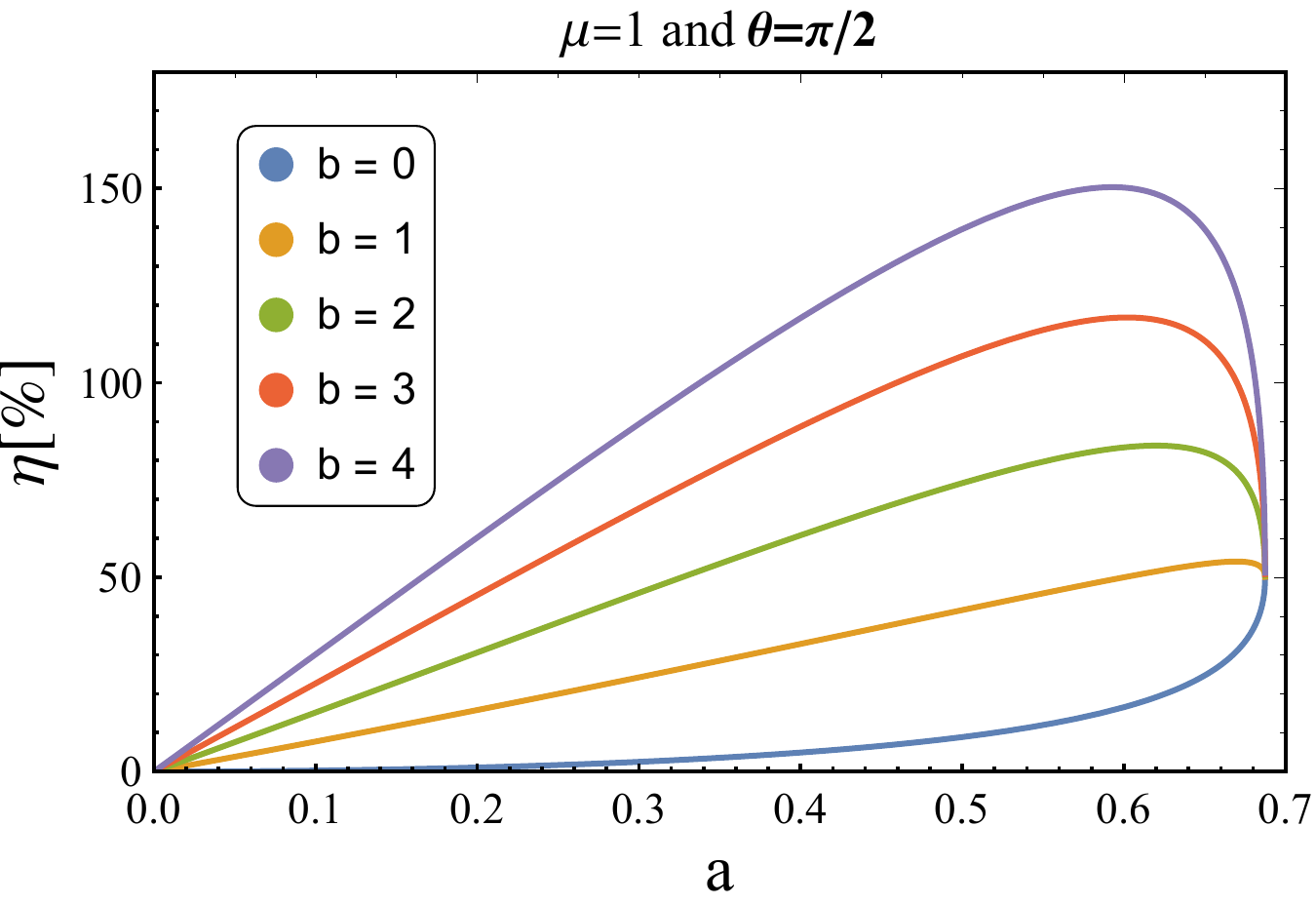}
\caption{Energy efficiency of MPP for MP rotating black holes having the maximum allowed $n$ rotations in $D=5$ (left column) and $D=6$ (right column) dimensions. 
One must keep in mind that here we have considered the case in which $a_1=a_2=a$ for both $D=5,\,6$. Note that the extremal value of the rotation parameter corresponds to $a_{ext}=0.5$ and $0.6873$ for $D=5,\,6$ dimensions, respectively.} \label{Fig1}
\end{figure*}
\begin{figure*}
\includegraphics[width=\columnwidth]{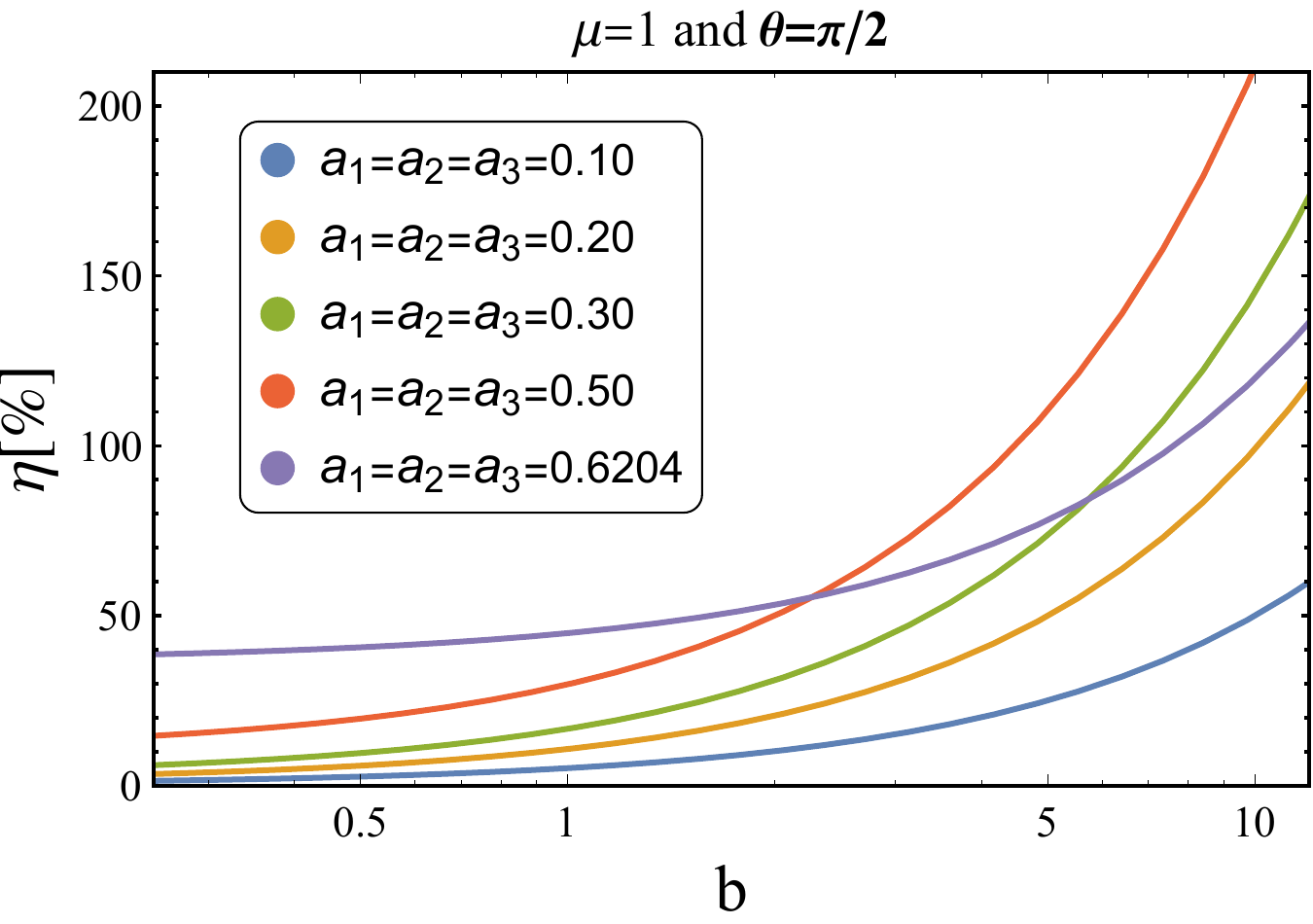}
\includegraphics[width=\columnwidth]{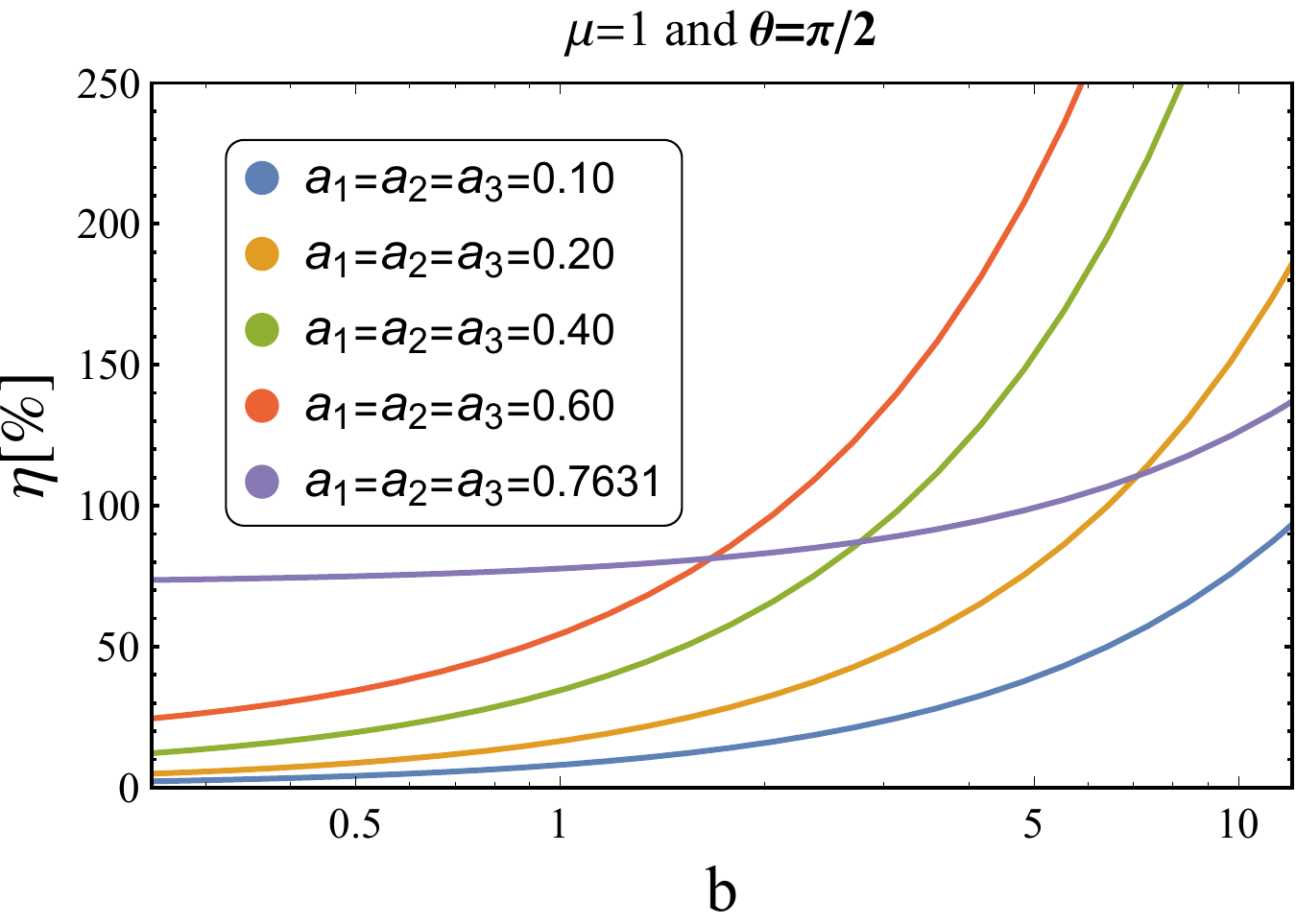}

\includegraphics[width=\columnwidth]{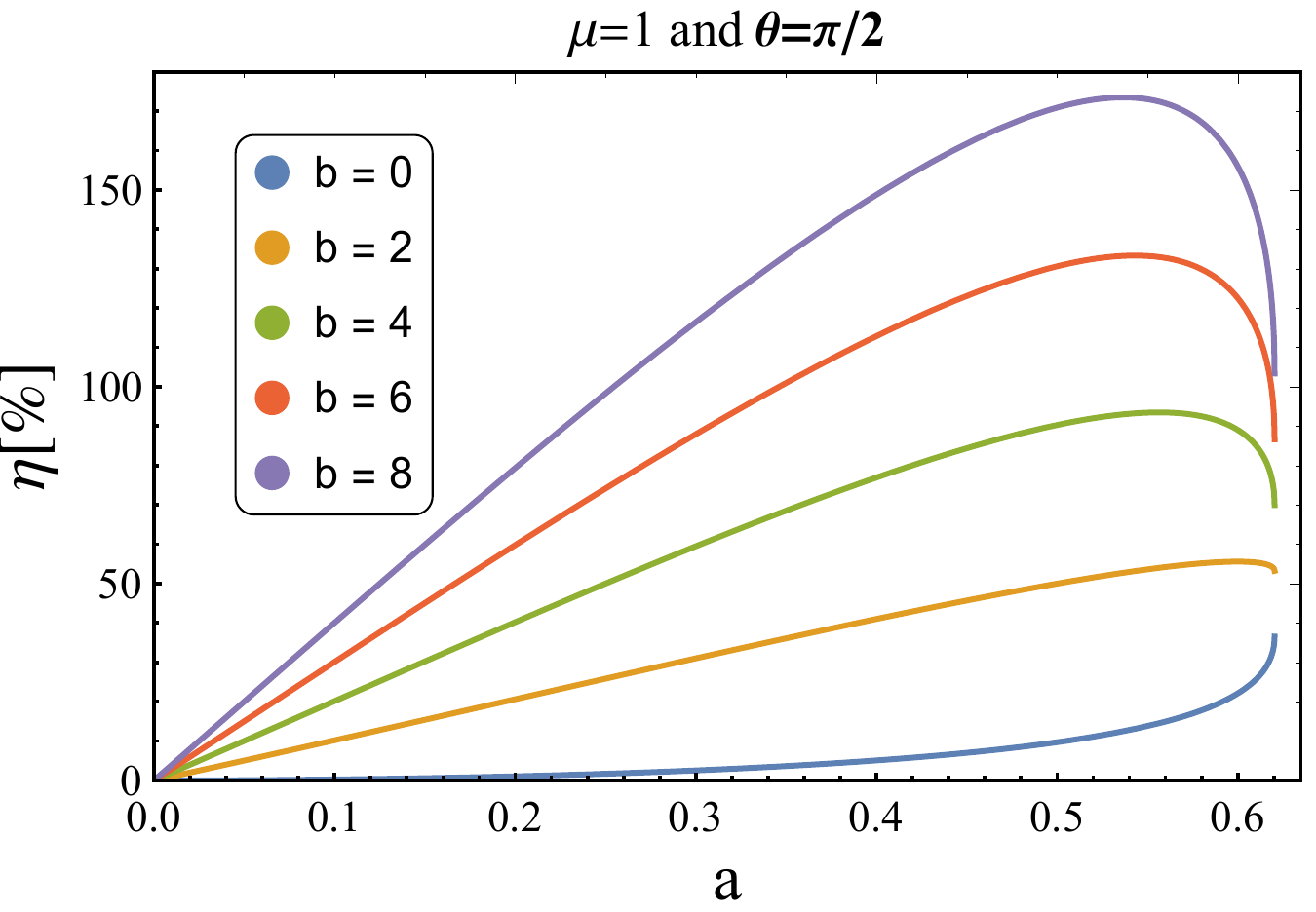}
\includegraphics[width=\columnwidth]{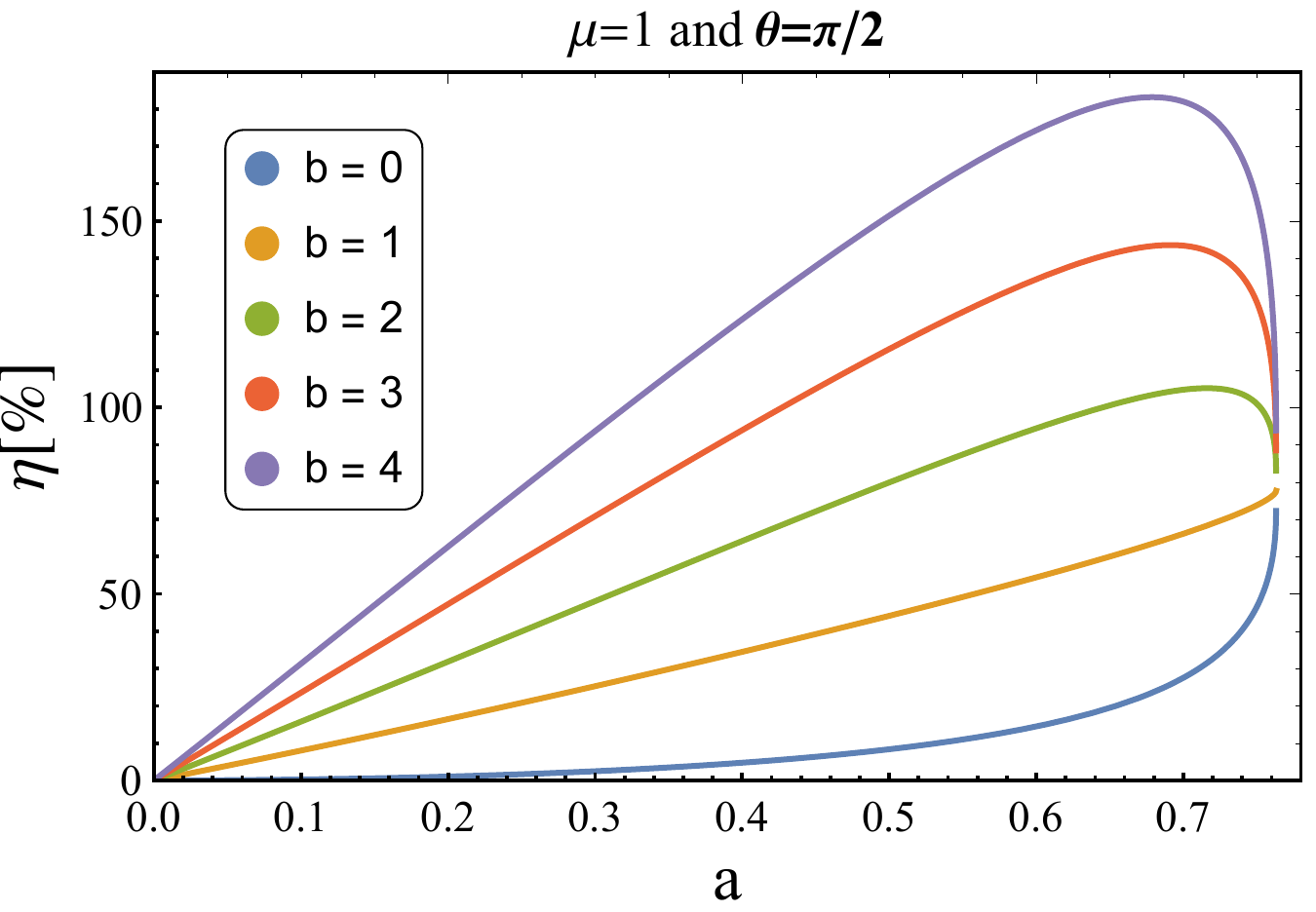}
\caption{Energy efficiency of MPP for MP rotating black holes having with the maximum allowed $n$ rotations in $D=7$ (left column) and $D=8$ (right column) dimensions. 
One must keep in mind that here we have considered the case in which $a_1=a_2=a_3=a$ for both $D=7,\,8$. Note that the extremal value of the rotation parameter corresponds to $a_{ext}=0.6204$ and $0.7631$ for $D=5,\,6$ dimensions, respectively.} \label{Fig2}
\end{figure*}

The energy efficiency depends on the splitting point, i.e.,  it reaches the maximum efficiency when the splitting point occurs very close to the horizon $r = r_{+}$ which will be discussed in the next section. For analysis we shall further consider $q/E_1 =q/m$ (i.e. $E_1/m_1=1$) at the splitting point. For further analysis of the energy efficiency expression we shall for simplicity set $a_1=a_2\cdots a_n=a$ and $\theta_1=\theta_2\cdots\theta_n=\theta=\pi/2$. With this in view, taking all $\Omega_{\phi_i}$ into account together with Eqs.~(\ref{Eq:D}) and (\ref{Eq.EM_pot}) we obtain the energy efficiency for $D=2n+1$ and $D=2n+2$ dimensions as    
\begin{align}\label{Eq:effMPP1}
    \eta^{(2n+1)}_{MPP}&=\frac{1}{2} \left(\sqrt{1+\frac{a^2}{r^2_{+}}}-1\right)\nonumber\\&+\frac{a}{\mu_{(2n+1)}}\,b_{(2n+1)}\left[1-\frac{n-1}{n}\frac{\mu r^2_{+}}{\Pi F}\right]\, ,
\end{align}
and
\begin{align}\label{Eq:effMPP2}
    \eta^{(2n+2)}_{MPP}&=\frac{1}{2} \left(\sqrt{1+\frac{a^2}{r^2_{+}}}-1\right) \nonumber\\&+\frac{a}{\mu_{(2n+2)}}\,b_{(2n+2)}\left[1-\frac{n-1}{n}\frac{\mu r_{+}}{2\Pi F}\right]\, ,
\end{align}
where we have denoted the magnetic field parameter $b$ for $D=2n+1, \,2n+2$, usually referred to as the interaction parameter as
\begin{eqnarray}\label{Eq:magnetic_par}
b_{2n+1}=\frac{qBG\mu_{(2n+1)}}{mc^4}\,\,\, \mbox{and}\,\,\,b_{2n+2}=\frac{qBG\mu_{(2n+2)}}{mc^4}\, .
\end{eqnarray} 
Here $B$ indicates the strength of the uniform external magnetic field. It is to be emphasized that the energy efficiency has two parts, one refers to the efficiency $\eta\vert_{q=0}$ for neutral particle while the other part $\eta\vert_{q\neq0}$ gives the MPP due to the magnetic field, as seen in Eqs.~(\ref{Eq:effMPP1}) and (\ref{Eq:effMPP2}).  

\section{The efficiency of energy extraction via MPP} \label{Sec:MPP2}

In this section we consider the energy efficiency via the well-established theoretical MPP mechanism that facilitates the energy extraction from a rotating black hole. It turns out that when one of rotations is switched off the efficiency of energy extraction would be arbitrarily large. For the efficiency of energy extraction we shall further examine $(n-1)$ and $n$ rotation cases separately.

\subsection{For $(n-1)$ rotations}

Interestingly, it turns out that higher dimensional black holes have only one horizon in the case when one of rotations is zero. This happens for both odd $D=2n+1$ and even $D=2n+2$ dimensions. Let's then recall the horizon equation, $\Delta$, Eq.~(\ref{eq:delta}) which for $(n-1)$ rotations yields 
\begin{eqnarray}\label{eq:delta1}
\frac{(r^2+a^2)...(r^2+a_{n-1}^2)}{r^{2(n-2)}}-\mu r^{5-D}=0 \, ,
\end{eqnarray}
which can simply be defined by the following form 
\begin{eqnarray}\label{eq:delta2}
r^{2(n-1)} &+& f_1(a_i) r^{2n-4} + ... - \mu r^{2n+1-D} \nonumber\\&+& a_1^2a_2^2 ...a_{n-1}^2 = 0 \, ,
\end{eqnarray}
with $f_{1}(a_{i})=a_1^2+a_2^2+...+a_{n-1}^2$.
To be more informative we further consider $D=5,\,7$ dimensional black holes with $(n-1)$ rotations. For $D=5,\,7$ the horizon equation Eq.~(\ref{eq:delta2}) gives implicitly $r_{+}$ as follows:
\begin{eqnarray}\label{eq:5dh}
r_{+}^2 = \left(\mu-a^2\right)\, ,
\end{eqnarray}
and 
\begin{eqnarray}\label{eq:6dh}
r_{+}^2 = -\frac{a_1^2+a_2^2}{2}+\frac{1}{2}\sqrt{4\mu+\left(a_1^2-a_2^2\right)^2 }\, ,
\end{eqnarray}
which for $a_1=a_2$ takes the form as $r_{+}^2 = \left(\sqrt{\mu}-a^2\right)$. It should be noted that $D=7$-dimensional black hole has three rotations (i.e., $a_1$, $a_2$ and $a_3$); see more details in Refs.~\cite{Shaymatov21a,Doukas11} addressing the cases in which one or more rotations are zero. As can be observed clearly from Eqs.~(\ref{eq:5dh}) and (\ref{eq:6dh}) the horizon equation Eq.~(\ref{eq:delta2}) can solve to give only one positive real root (i.e., only one horizon) for both $D=2n+1, 2n+2$, thus resulting in occurring no extremality condition which can further give rise to infinitely large energy efficiency even via PP as well as MPP. Let us then turn to obtain the efficiency of energy extraction for $(n-1)$ rotations, for example, Eqs.~(\ref{Eq:effMPP1}) and (\ref{Eq:effMPP2}) for $D=5,\,7$ will respectively reads as 
\begin{eqnarray}
    \eta_{q=0}|_{a_1=a;\,a_2=0}&\sim \left(\frac{1}{\mu-a^2}\right)^{(1/2)}\, ,
\end{eqnarray}
and 
\begin{eqnarray}
    \eta_{q=0}|_{a_1=a_2=a;\, a_3=0}&\sim \left(\frac{1}{\sqrt{\mu}-a^2}\right)^{(1/2)}\, .
\end{eqnarray}
This clearly shows that the energy efficiency can yield arbitrarily large values in the limit of $a\to\mu$ even for PP (i.e., $b=0$). This happens because the rotation or angular momentum is not bounded from above due to the existence of no extremality condition when considering black holes having $(n-1)$ rotations. Similar result in relation to infinitely large energy efficiency when one rotation is infinitesimally small (i.e., $a_1\gg a_2$) has also been noticed in Ref.~\cite{Nozawa05}. Similarly, one can observe the same result for even $2n+2$ dimensions. Therefore, the energy efficiency even without MPP can approach infinitely large values for higher dimensional $D>4$ black holes with $(n-1)$ rotations regardless of dimension being odd $D=2n+1$ or even $D=2n+2$. It is to be emphasized that the inclusion of MPP (i.e., $b\neq 0$) here can only support and enhance the efficiency of energy extraction from MP black holes.

\subsection{For $n$-rotations}

As shown earlier black hole with $(n-1)$ rotations has only one horizon, thereby it comes out in favor of infinitely large energy efficiency for both $D=2n+1$ and $D=2n+2$ dimensions. Hence, the black holes can be characteristically different from the ones with $n$ rotations in $D>4$ dimensions. For $n$-rotations the horizon equation, $\Delta=0$, is given by 
\begin{eqnarray}\label{Eq:delta}  
r^{2n} + f_1(a_i) r^{2n-2} + ... - \mu r^{3-D+2n} + a_1^2a_2^2 ...a_n^2 = 0\, ,\nonumber\\
\end{eqnarray}
which always has two positive roots for which black hole can exist two horizons accordingly despite the dimension being odd $D=2n+1$ or even $D=2n+2$ \cite{Shaymatov21a,Shaymatov23Univ}. For example, $r_{\pm}$ for $D=5$ reads as follows \cite{Shaymatov19a}  
\begin{eqnarray}
r_{\pm}^2 = \frac{\mu-a_1^2-a_2^2}{2}\pm \sqrt{\frac{\left(\mu-a_1^2+a_2^2\right)^2}{4}- a_1^2 a_2^2}\, ,
\end{eqnarray}
which for $a_1=a_2=a$ yields 
$$r_{\pm}^2=\frac{\mu}{2}- a^2\pm\frac{1}{2}\sqrt{\mu \left(\mu-4 a^2\right)}\, .$$ 
For that the necessary condition for extremality is then satisfied well, i.e., $\mu=4a^2$. This therefore gives an allowed range of rotation; i.e. it is bounded from above, $a\leq \sqrt{\mu}/2$. In addition, the approach $a\to \sqrt{\mu}/2$ allows the splitting point of incident particle to occur around black hole, especially very close to the horizon  $r_{st}<r\leq r_{+}$ where the efficiency of energy extraction reaches its maximum. The question then arises, could the efficiency rate of energy extraction be the same with the one for $(n-1)$ rotations? The answer however comes out to be no, i.e., the efficiency that we further show explicitly cannot become arbitrarily large for the maximum allowed $n$ rotations in the case of the Penrose process. To this end, we shall consider the MPP and examine the efficiency of energy extraction process. This is what we intend to address and gain a deeper understanding in the following.
 \begin{table}
\begin{center}
\caption{The threshold values of the magnetic parameter $b$ required for the efficiency of energy extraction to exceed $>100\%$ for various values of the rotation parameter. Note that here we have considered the case in which $a_1=a_2=a$ for $D=5,\,6$ and $a_1=a_2=a_3=a$ for $D=7,\,8$, respectively. }\label{table1}
\resizebox{.38\textwidth}{!}
{
\begin{tabular}{c c| c c | c c }
 \hline \hline
\cline{3-4}\cline{5-6}
& $a$    & $D=5$ & $D=6$     &$D=7$ & $D=8$     \\
\hline
& 0.1   & 20.1548 & 13.3447   & 20.0509 & 12.8544     \\
& 0.2   & 10.3428 & 6.69123    & 10.1076 & 6.42317   \\
& 0.3     & 7.29715 & 4.48663   & 6.84556 & 4.27790   \\
& 0.4       & 6.27322 & 3.40308 & 5.28343  & 3.20470   \\
& 0.5       & $-$ & 2.78938 & 4.48119  & 2.56189   \\
& 0.6       & $-$ & 2.49669 & 4.65416  & 2.13956   \\
& 0.7       & $-$ & $-$ & $-$  & 1.87652   \\
 \hline \hline
\end{tabular}
}
\end{center}
\end{table}

Let us then turn to the analysis of the energy efficiency for MP black holes with $n$-rotations. We demonstrate the energy efficiency driven out through MPP at the splitting point occurring in the close vicinity of the horizon in Figs.~\ref{Fig1} and~\ref{Fig2}.  Fig.~\ref{Fig1} depicts the profile of the efficiency of energy extraction from black holes in $D=5,\,6$ as a function of the magnetic field parameter $b$ and the rotation parameter $a$, 
while Fig.~\ref{Fig2} shows similar behavior of the efficiency in $D=7,\,8$ for various combinations of rotation parameters (i.e., $n=2,\, 3$ for $D=5,\,6$ and $D=7,\,8$) and the magnetic field parameter. As can be observed from Eqs.~(\ref{Eq:effMPP1}) and (\ref{Eq:effMPP2}) as well as from Fig.~\ref{Fig1} the energy efficiency respectively reaches the maximum of  
\begin{eqnarray}\label{eff_pp1}
    \eta|^{D=5}(a\to a_{ext})&=\frac{1}{2} \left(\sqrt{1+\frac{a^2}{r^2_{+}}}-1\right)\sim20.7\%\, ,\\
    \eta|^{D=6}(a\to a_{ext})&=\frac{1}{2} \left(\sqrt{1+\frac{a^2}{r^2_{+}}}-1\right)\sim 50\%\, ,
\end{eqnarray}
when considering $b=0$. 
However, the MPP part also goes to $\eta\vert_{q\neq0}= 0$ as $a\to a_{ext}$ for rotation, as shown in the top and bottom rows of Fig.~\ref{Fig1}, similarly to what was observed in the rotating Kerr case in $D=4$ dimensions \cite{Dadhich18mnras}. Here we note that the horizon for the above expression respectively takes values as $r_{+}(a_{ext})=0.5,\,\sim 0.3969$ in $D=5,\,6$, where mass parameter has been set $\mu=1$. 
 However, the energy efficiency is enhanced drastically when the interaction of a magnetic field with a charged particle is included, referred to MPP. Fig.~\ref{Fig1} (top row) clearly shows that the energy efficiency curves shift towards up to its higher values with the increase in the value of $a$. This happens because the splitting point gets closer to the black hole's horizon whenever $a$ approaches its larger values. Another interesting point to note here is that the energy efficiency slightly increases with increasing the parameter $b$ for both $D=5,\,6$ cases. To achieve the maximum efficiency, it requires $b>1$ for the dimensionless magnetic parameter values, thus allowing the magnetic field effect to dominate over gravity all through. As can be seen from Fig.~\ref{Fig1} (bottom row) that the energy efficiency curves shift upward toward its larger values and go over 100$\%$ as a consequence of an increase in the value of the magnetic parameter $b$ for both $D=5,\,6$ cases. 
 It then emerges that the MPP can eventually become more efficient than the PP due to the fact that the magnetic field circumvents the formidable velocity threshold of PP, causing the efficiency of energy extraction to increase enormously. To understand more quantitatively and astrophysically, we need to estimate the constraint value of the dimensionless parameter $b$ for which the efficiency exceeds 100$\%$. For that we will impose the condition $\eta \geq 100\%$ as given in Eqs.~(\ref{Eq:effMPP1}) and (\ref{Eq:effMPP2}). Consequently, it solves to give $b \geq 6.27$ and $3.40$, respectively, for a charged particle around $D=5,\,6$ dimensional black holes with the same rotations, for example, $a_1=a_2=0.4$. Furthermore, we provide a more detailed numerical analysis and tabulate the threshold values of the magnetic parameter $b$ required for the efficiency of energy extraction to exceed $100\%$ in Table~\ref{table1}. This constrain value of $b$ can be expected to be smaller than these values for the same rotations in $D=7,\,8$ dimension cases. As can be observed from Fig.~\ref{Fig1} that the MPP can enhance the energy efficiency that can be likely to grow significantly and thus reaches up its possible maximum values for $D=5,\, 6$, which is surprisingly quite large and exceeding over 100$\%$.

Similarly, we begin to examine the maximum efficiency of PP for $D=7,\,8$. Hence, we can now be somewhat more quantitative. That is, the efficiency does respectively take 
\begin{eqnarray}\label{eff_pp1}
    \eta|^{D=7}(a\to a_{ext})&=\frac{1}{2} \left(\sqrt{1+\frac{a^2}{r^2_{+}}}-1\right)\sim36.5\%\, ,\\
    \eta|^{D=8}(a\to a_{ext})&=\frac{1}{2} \left(\sqrt{1+\frac{a^2}{r^2_{+}}}-1\right)\sim 72\%\, ,
\end{eqnarray}
when considering $b=0$ for $a_{ext}$ in $D=7,\,8$ dimensions. 
Interestingly, we observe from Fig.~\ref{Fig2} that the energy efficiency remains above $100\%$ when $a\to a_{ext}$ in both $D=7,\,8$ dimensions. From the bottom row of Fig.~\ref{Fig2}, it is clearly seen that $\eta|_{b\neq 0}>\eta|^{D=7,\,8}(a\to a_{ext})$ is always satisfied as a consequence of the presence of $b$. This is the remarkable difference from $D=5,6$ dimensions, where MPP part goes to zero, as shown in both top and bottom rows of Fig.~\ref{Fig1}. However, the energy efficiency rate can rapidly increase once MPP effect is included. Hence, its growing rate is much prominent due to the MPP, similarly to what is observed in $D=5,\, 6$ dimensions. However, it must be noted that the increasing rate of the efficiency through the MPP grows significantly as the number of rotations increases. The efficiency of energy extraction is greatly enhanced in both $D=7,\,8$ dimensions, eventually exceeding more than 100$\%$ as $b$ increases for appropriate values of rotation parameter. This can be seen in both the top and bottom rows of Fig.~\ref{Fig2}. We also estimate the constraint values of $b$ for the charged test particle. Similarly, by imposing the condition $\eta \geq 100\%$ we consequently find $b \geq 5.28$ and $3.20$ for $D=7,\,8$ with the same $a_1=a_2=a_3=0.4$ (see details, e.g., in Table~\ref{table1}). From Fig.~\ref{Fig1} and \ref{Fig2}, one can observe that the energy efficiency via the MPP can reach up to its huge values in all cases. This might happen due to the dimensionless parameter $b$ that can take any value astrophysically in the case of elementary particles. Here, the magnetic field effect with the rotation plays an important role as a powerful tool in increasing the efficiency of energy extraction exceeding over $>100\%$. Therefore, the rotational energy of MP black holes in all higher $D>4$ can be efficiently driven out through the MPP as stated by the escaping charged particle interacting with the magnetic field existing in the environment surrounding black hole.

\section{Conclusions}\label{Sec:Conclusion}
It has widely been believed that astrophysical rotating black holes are regarded as the primary sources of energy for highly powerful astrophysical phenomena. Hence, black hole energetics is increasingly important to gain a deeper understanding about their rotational energy. With this in view, higher dimensional MP black holes would also be an alternative source of information about high energy astrophysical objects and strong gravity field regimes, together with the surrounding geometry. Therefore, it is important to test the effects that stem from higher dimensions on black hole energetics. In this paper, we extensively studied the energy efficiency of MPP for odd $D=2n+1$ and even $D=2n+2$ dimensional black holes with both $(n-1)$ and $n$ rotations.  

We have shown that the efficiency of energy extraction can yield infinitely large values in the case when the black hole has only rotation (i.e., $a_2=0$) in both $D=5,\,6$ dimensions. This large amount of energy efficiency can also be reached even by PP (i.e., $b=0$). This happens because there exists no extremality (i.e., only one horizon), thereby leading to the case in which rotation is not constrained from above. This is the case for all higher dimensional black holes with $(n-1)$ rotations, as there is no extremality condition irrespective of whether the dimension is odd $D=2n+1$ or even $D=2n+2$. Therefore, the energy efficiency can approach infinitely large values. This result has also been noticed in Ref.~\cite{Nozawa05}. This is a characteristic distinction in contrast to the cases with $n$ rotations in all $D>4$ dimensions. Additionally, the contribution of MPP here can more likely enhance the efficiency of energy extraction.      

However, it has been shown that the energy efficiency is not expected to be the same as in the case of $(n-1)$ rotations, thus referring to limited energy efficiency for the maximum allowed $n$ rotations; i.e., $\eta$ $\sim$ 20.7$\%$ and 50$\%$ for $D=5,\,6$ while $\sim$ 36.5$\%$ and 72$\%$ for $D=7,\,8$ dimensions when considering $b=0$ and $a\to a_{ext}$, referred to as pure PP. In fact, one can observe that the horizon equation, $\Delta=0$, has only two positive roots, thereby leading to the existence of two horizons of black holes with $n$ rotations in both $D=2n+1$ and $D=2n+2$ dimensions \cite{Shaymatov21a,Shaymatov23Univ}. Therefore, the energy efficiency for PP cannot become arbitrarily large but can be limited instead for the maximum allowed $n$ rotations unless there exists another promising thought mechanism. To that end, we further considered a well-established mechanism, referred to as MPP, and showed its impact on the efficiency of energy extraction through the interaction of magnetic field with the escaping charged particle.

We have demonstrated that energy efficiency increases and tends to reach higher values as a consequence of the increase in the magnetic parameter $b$ for both $D=5,\,6$ dimensions. It should be noted that it requires overestimated values of the parameter $b$ for high energy efficiency extracted from the black hole. Additionally, with the rise in the value of the rotation parameter $a$, we have shown that the efficiency of energy extraction is significantly enhanced by the MPP, and thus it begins to grow and goes over $>100\%$ for $D=5,\, 6$ dimensions, respectively. This enhancement occurs due to the splitting point near the black hole's close vicinity, especially very close to its horizon $r_{+}$, as depicted in Fig.~\ref{Fig1}. To provide a quantitative understanding, we have determined the constraint values of $b$ that allow the efficiency to exceed 100$\%$, finding them to be approximately $b \geq 6.27,\,3.40$ for $D=5,\,6$ in the presence of a charged particle around MP black hole with rotation $a=0.4$. Similarly, for $D=7,\,8$ the values are $\geq 5.28,\,3.20$, respectively, with the same rotation; see details, e.g., in Table~\ref{table1}. Furthermore, we have shown that the efficiency of energy extraction for $D=7,\,8$ is also strongly enhanced by the MPP, similar to the cases of $D=5,\, 6$. The rate of growth in energy efficiency is much prominent in $D=7,\,8$ dimensions, increasing significantly due to the MPP. To be somewhat more quantitative with estimates of energy efficiency, we have demonstrated that it can easily exceed over $>100\%$ in $D=7,\,8$ dimensions, as shown in Fig.~\ref{Fig2}. Interestingly, we observed that, for the corresponding values of $b$ the energy efficiency remains above $100\%$ as $a\to a_{ext}$ in both $D=7,\,8$, unlike in $D=5,6$ where the MPP part goes to zero (see, for example, Fig.~\ref{Fig1}). It has been established that the efficiency of energy extraction can surpass $100\%$ in all higher dimensions (i.e., $D>4$) through the MPP.

It is remarkable to note that the energetics of higher dimensional MP black holes were studied by several authors from different perspectives \cite{Nozawa05,Prabhu10,Abdujabbarov13bsw}. For example, the energy extraction from higher dimensional MP black holes and black rings was considered with the usage of the PP in Ref.~\cite{Nozawa05}. Later, the center of mass energy via the BSW (Banados-Silk-West) mechanism in the vicinity of a $D=5$ dimensional MP black hole with two rotation parameters was studied in Ref.~\cite{Abdujabbarov13bsw}, showing that the head-on collision energy of two particles can be arbitrarily high near an extremal black hole. Also, the energetics of a rotating charged black hole in $D=5$ dimensional supergravity was studied in Ref.~\cite{Prabhu10}, addressing the energy extraction from the black hole using the PP. In all of this work, the magnetic field effect on the extraction of energy efficiency was not included. From an astrophysical viewpoint, it is increasingly important to explore the pivotal role of the magnetic field in altering the geodesics of charged test particles in the close vicinity of a black hole. Based on the results presented in this paper, MPP could potentially be a more effective and viable mechanism for extracting the rotational energy of higher dimensional black holes with $n$ rotations, significantly enhancing the efficiency of energy extraction. This would be a primary astrophysical significance, as it does not exclude the existence of higher dimensional astrophysical black holes that could serve as sources of highly energetic astrophysical phenomena and play a crucial role in explaining these extremely powerful events.

\acknowledgments

We are grateful to the anonymous referees for their insightful comments
and constructive suggestions that definitely helped us improve the
clarity and quality of the manuscript. This work is supported by the National Natural Science Foundation of China under Grant No. 11675143 and the National Key Research and Development Program of China under Grant No. 2020YFC2201503.

\appendix

\bibliographystyle{apsrev4-1}  
\bibliography{Ref,gravreferences}

\end{document}